\documentclass[a4paper]{article}
\usepackage{amscd,amsmath,amssymb}
\usepackage[usenames, dvipsnames]{color}

\newcommand{\mF}{\mathfrak F}

\newcommand{\p}[1]{(\ref{#1})}

\newcommand{\cC}{{\cal C}}

\newcommand{\cZ}{{\cal Z}}

\newcommand{\cU}{{\cal U}}

\newcommand{\cA}{{\cal A}}

\newcommand{\cJ}{{\cal J}}
\newcommand{\cX}{{\cal X}}
\newcommand{\cW}{{\cal W}}
\newcommand{\cY}{{\cal Y}}
\newcommand{\cV}{{\cal V}}

\newcommand{\cN}{{\cal N}}
\newcommand{\hJ}{{\hat J}}

\def\hX{\widehat{X}}
\def\hZ{\widehat{Z}}
\def\hU{\widehat{U}}
\def\hW{\widehat{W}}
\def\hY{\widehat{Y}}
\def\hJ{\widehat{J}}
\def\hV{\widehat{V}}

\newcommand{\be}{\begin{equation}}
\newcommand{\ee}{\end{equation}}
\newcommand{\bea}{\begin{eqnarray}}
\newcommand{\eea}{\end{eqnarray}}

\newcommand{\ba}{\begin{array}} \newcommand{\ea}{\end{array}}

\def\im{{\rm i}}

\newcommand{\nn}{\nonumber}
\def\theequation{\arabic{section}.\arabic{equation}}

\begin{document}
\begin{flushright}
	\end{flushright}\vspace{1cm}
\begin{center}
	{\Large\bf  $\cN=8$ superconformal mechanics: direct construction}
\end{center}
\vspace{1cm}

\begin{center}
{\Large\bf Sergey Krivonos${}^{a}$ and Armen Nersessian${}^{b,c,a}$}
\end{center}

\vspace{0.2cm}

\begin{center}
	
	\vspace{0.3cm}
	
	${}^a${\it
		Bogoliubov  Laboratory of Theoretical Physics, JINR,
		141980 Dubna, Russia}
				
	${}^b$ {\it Yerevan Physics Institute,
			2 Alikhanyan Brothers  St., Yerevan, 0036, Armenia}
			
	${}^c$ {\it Institute of Radiophysics and Electronics, Ashtarak-2, 0203, Armenia }
	
	\vspace{0.5cm}
	
{\tt   krivonos@theor.jinr.ru,arnerses@yerphi.am}
\end{center}
\vspace{2cm}

\begin{abstract}
	\noindent In the present paper we constructed the supercharges and Hamiltonians for all variants of superconformal mechanics associated with the superalgebras $osp(8|2), \mF(4), osp(4^\star |4)$, and $su(1,1|4)$. The fermionic and bosonic fields involved were arranged into generators spanning $so(8), so(7), so(5)\oplus su(2)$ and $su(4) \oplus u(1)$ $R$-symmetry currents of the  corresponding superconformal algebras. The bosonic and fermionic parts of these  $R$-symmetry generators   separately define the constants of motion and form the same  algebras. 	The  angular part of the supercharges defining the system have the structure 	``$(R-symmetry\; generators)\, \times\, fermions$'' while the angular part of Hamiltonian is just a proper sum of full  Casimir operators and its 	purely bosonic and fermionic parts.
	We also constructed the explicit embedding of  the algebras $so(7), \, so(5) \times su(2)$, and $su(4)\times u(1)$ into $so(8)$, which provide the possibility to explicitly construct the corresponding supercharges.
\end{abstract}
\vskip 1cm
\noindent

\vskip 1.0cm


\newpage

\pagenumbering{arabic}
\setcounter{page}{1}
\section{Introduction}
The superconformal mechanics play an important role as the simplest examples 
in studying systems with supersymmetry in higher space-time dimensions, involving the
proper supersymmetrized  gauge, gravitation, and cosmological theories. They are also
directly related to diverse superextensions of d= 1 integrable systems. The most advanced and suggestive method to deal with supersymmetric theories is the superfield approach, which, however,
works  perfectly in the case of lower  $\cN \leq 4$ supersymmetries but acquires a lot of complications for theories with $\cN=8$ supersymmetry. The main reason is the large dimension of the measure of the corresponding $\cN=8$ superspace, which prevents us from immediate writing   superfield actions. The commonly adopted way to deal with $\cN=8$ supersymmetric theories is to construct the action in terms of several $\cN=4$ superfields. Then, one has to check the existence of another implicit $\cN=4$ supersymmetry which together with the explicit supersymmetry could  form $\cN=8$ supersymmetry \cite{VP}-\cite{toppan2}. Evidently, such a construction of  $\cN=8$  supersymmetric systems is much more complicated than $\cN=4$ supersymmetric ones because the number and content of $\cN=4$ superfields involved is {\it a priori} unknown. Moreover, in some cases, to have an appropriate component action, one has to use the so-called oxidation procedure to avoid the second-order Lagrangian for fermions. 

Despite all these troubles, several important steps forward were made in a series of papers \cite{ABC,Tolik,Jim0,Jim,toppan1,toppan2} where all $\cN=8$ supermultiplets were classified and their $\cN=4$ content was clarified. Thus, everything was ready and the search for $\cN=8$ superconformal mechanics was started. The line of the story looks like:
\begin{itemize}
	\item Prior to the clarification of  $\cN=8$ supermultiplets, the $\cN$-extended superconformal mechanics based on the algebra $su(1,1| \cN/2)$ was constructed in \cite{leva}.
	The basic supermultiplet in the $\cN=8$ case was $(1,8,7)$, so the Hamiltonian contained 8 fermions and the single bosonic field, dilaton. Later on, in \cite{toppan1}	the most general $\cN=8$ invariant action was constructed based on the  same  $\mathbf{(1,8,7)}$ supermultiplet.  It depends on two real parameters: $a$ and $b$ and gives an interaction term; by setting $a=0$, one recovers the action of ref. \cite{leva};
	\item In \cite{Diac}, the superconformal $N=8, d=1$ action was constructed for the low
	energy effective dynamics of a D0-brane moving in D4-brane and/or orientifold plane
	backgrounds (see also \cite{BMZ});
	\item Next, in \cite{BIKL}  new models of $\cN=8$ superconformal mechanics associated with the off-shell $\cN=8, d=1$ supermultiplets $(3,8,5)$ and $(5,8,3)$ were constructed. These two multiplets are derived as $\cN=8$ Goldstone superfields and correspond to nonlinear realizations of the $\cN=8, d=1$ superconformal group $OSp(4^\star |4)$ in its supercosets $OSp(4^\star |4)/U(1)_R \times SO(5)$ and $OSp(4^\star |4)/SU(2)_R \times SO(4)$, respectively; 
	\item  In 2007, D.~Delduc and E.~Ivanov   constructed the $\mF(4)$ superconformal action for the $(1,8,7)$ supermultiplet \cite{F4scm}.  Later on, it was shown in \cite{toppan2} 
	that this $\mF(4)$ superconformal action is recovered
	by setting $b=0$ in the action in  \cite{toppan1}. Finally in 2018, the quantum superconformal Hamiltonian of the $\mF(4)$ model was constructed in \cite{toppan3}. In addition, in  \cite{fed2}, a multicomponent version of $\mF(4)$ superconformal mechanics was constructed;
	\item Recently, S.~Fedoruk and E.~Ivanov have presented the new model of $\mathcal{N}=8$ supersymmetric mechanics constructed by using  the so-called   dynamical and semi-dynamical $\mathcal{N}=4$ multiplets \cite{fed1}.
	
	Later on, in  \cite{hkn1}, we showed that this system admits dynamical  $OSp(8|2)$ superconformal symmetry and describes a specific $\cN=8$ supersymmetric extension of the free particle on the {\sl eight-dimensional cone } embedded in nine-dimensional pseudo-Euclidean space with the fermionic part that can be interpreted as a spin-orbit coupling term.
\end{itemize}

One should note that the final on-shell component actions with excluded auxiliary fields are highly indirectly related to  the superfield Lagrangians the story was started with. The existence of additional, say superconformal symmetry, does not help much. Indeed, in contrast with the $\cN=4$ case with mainly one superconformal algebra $D(1,2;\alpha)$, in the $\cN=8$ supersymmetric case, we have four different superconformal algebras $osp(8|2), \, \mF(4), \, osp(4^\star |4)$, and $su(1,1|4)$ (see e.g. \cite{VP}).
In contrast, the component approach, especially its Hamiltonian version, looks quite promising to
deal with. The main preferences of the component approach to $\cN=8$ superconformal mechanics can be summarized as follows:
\begin{itemize}
	\item In all variants of the superconformal mechanics with a different number of supersymmetries,
	the superconformal symmetry is spontaneously broken, as the result of which the dilaton $u$ plays the master role in the form of supercharges and Hamiltonian. Roughly speaking, the supercharges have the form
	\be\label{01} Q \sim p_r \psi + \frac{1}{r} \left[ R-symmetry \; generators\right] \psi .\ee
	Here, $r=e^{u/2}$ and $p_r$ is the corresponding momentum.
	\item The explanation of this form \p{01} is quite simple: due to the presence of the dilaton
	fields, one can always realize the dilaton $D$, conformal boost $K$ and the conformal supersymmtey $S$  generators as
	\be\label{02}
	{\cal D}= \frac{1}{2} r p_r, \quad {\cal K}= \frac{1}{2}r^2, \qquad S =  r \psi .
	\ee
	\item Thus, the form of the supercharges \p{01} is unique  to produce the main relations of any superconformal algebra
	\be
	\{Q,S\} \sim {\cal D} + \left[ R-symmetry \; generators\right].
	\ee
\end{itemize}

The Hamiltonian analysis of the $\cN=8$ supersymmetric mechanics \cite{fed1} performed in  \cite{hkn1} confirms the properties we just discussed and revealed several new peculiarities of the model:
\begin{itemize}
	\item The system possesses dynamical  $OSp(8|2)$ superconformal symmetry;
	\item The components of semi-dynamical supermultiplets can be treated as  coordinate-momenta
	variables in the first order formulation of the system. From this point of view, the initial  interpretation of  the model as a $\mathcal{N}=8$ supersymmetric extension of a  one-dimensional isospin particle turned into superconformal mechanics with eight bosonic and
	eight fermionic physical components;
	\item The fermionic and bosonic fields involved can be arranged into generators spanning $so(8)$
	$R$-symmetry of the  $osp(8|2)$ superalgebra. The bosonic and fermionic parts of these  $so(8)$-generators   separately define the constants of motion and form the same $so(8)$ algebra;
	\item The  supercharges defining the system indeed have the structure visualated in \p{01}.
	\	Correspondingly, the constructed  Hamiltonians read 
    \be\label{03}
  {\cal H} =\frac{1}{2}p_r^2 +\beta \frac{\cC}{r^2}. 
  \ee
	Here $\beta$ is the constant, and $\cC$ denotes the Casimir operators of the $R$-symmetries of the system constructed from bosonic, fermionic and bosonic+fermionic components.  
	\item The system describes the specific $\mathcal{N}=8$ supersymmetric extension of a free particle on the  {\sl eight-dimensional cone } embedded in nine-dimensional pseudo-Euclidean space with the fermionic part that can be interpreted as a spin-orbit coupling term.
\end{itemize}

In this paper,  we give the Hamiltonian formulation of the systems with all possible superconformal symmetries. The idea of construction comes from  $OSp(8|2)$ superconformal mechanics \cite{fed1, hkn1}. It consists in the above-mentioned special form of  supercharges which are constructed with the help of the generators of $R$-symmetries. The generators of $R$-symmetries are constructed from
eight fermions $\{ \phi^{i A }, \chi^{a \alpha}; \quad i,A,a, \alpha =1,2\}$ and sixteen bosonic variables $\{B^{a A}_m, Z^{i \alpha}_m;\quad m=1,2\}$. The $so(8)$ generators of bosonic and fermionic realizations are defined in  the Appendix, \p{bosonicso8}, \p{fsu2}, \p{fV}.

The key peculiarity of the  algebra $so(8)$ is its representation in terms
of four commuting $su(2)$ algebras with the  generators ($\cJ^{i j}, \cW^{A B}, \cX^{a b}, \cY^{\alpha \beta}$) and generators from the coset $so(8)/su(2)^4\, - \,\cV^{i A a \alpha}$ obeying the brackets
$$
 \left\{\cV^{i A a \alpha},\cV^{j B b \beta}\right\} = -4 \left( \epsilon^{AB}\epsilon^{ab} \epsilon^{\alpha \beta} \cJ^{ij} + \epsilon^{ij}\epsilon^{ab} \epsilon^{\alpha \beta} \cW^{AB}+ \epsilon^{ij}\epsilon^{AB} \epsilon^{\alpha \beta} \cX^{ab} + \epsilon^{ij}\epsilon^{AB} \epsilon^{a b} \cY^{\alpha \beta}\right). 
$$
Such representation of $so(8)$ algebra gives the easiest way to pass to  smaller $R$-symmetry
algebras of  other $\cN=8$ superconformal algebras:
\begin{itemize}
	\item The $so(7)$  $R$-symmetry algebra of $\mF(4)$ superconformal algebra can be defined by the generators
	$$ so(7) \propto \{\cJ^{i j}+\cY^{ij},\, \cW^{A B},\, \cX^{a b},\, \cV^{i A a j}+\cV^{j A a i}\};$$
	\item The $so(5)$  $R$-symmetry algebra of $Osp(4^\star |4)$ superconformal algebra can be defined by the generators
	$$ so(5) \propto \{ \cW^{A B},\, \cX^{a b},\, \cU^{A a} = \epsilon_{ij} \cV^{i A a j}\}.$$
	The corresponding $su(2)$  $R$-symmetry algebra commuting with $so(5)$ algebra is spanned by the generators 
	$$ su(2) \propto \{\cJ^{i j}+\cY^{ij}\};$$ 
	\item The $su(4)$  $R$-symmetry algebra of $su(1,1|4)$ superconformal algebra is spanned by the generators
	$$su(4) \propto \{\cJ^{ij}+\cY^{ij},\,\cX^{ab}+\cW^{ab},\,\cZ^{i\,a\,b\,j} \equiv \cV^{i\,a\,b\,j}+\cV^{i\,b\,a\,j}+\cV^{j\,a\,b\,i}+\cV^{j\,b\,a\,i}\}.$$
	Correspondingly, the $u(1)$ generators commuting with $su(4)$ algebra and presented in $su(1,1|4)$ superconformal algebra can be defined as 
	$$U(1) \propto \{\cU = \cU^i{}_i = \cV^{i\,a}{}_{a\,i} \}.$$
\end{itemize}  

Using these relations , in  {\sl Sections 3-5} we construct the supercharges and the Hamiltonians for all $\cN=8$ superconformal mechanics. All supercharges were constructed as  in  \p{01}, and  all our Hamiltonians have the structure \p{03}.

In  {\sl Section 6} we show that the bosonic parts of the constructed models describe the free particles on the cones embedded in the corresponding  (pseudo-)Euclidean spaces, while the fermionic parts can be interpreted as  the spin-orbital interaction terms.

In {\sl Conclusion} we review the obtained results and mention possible future developments.

In {\sl Appendix} we present the definitions of our bosonic and fermionic fields and 
the explicit expressions for algebra $so(8)$ written in the notation used.

\section{$\cN=8$ superconformal algebra and Ansatz for the supercharges}

In what follows we will need the generators that extend $\mathcal{N}=8$ Poincar\'{e} supersymmetry
\be\label{N8P}
\left\{Q^{ i\,A} , Q^{j\, B} \right\} = 4 \imath\epsilon^{ij} \epsilon^{AB} {\cal H}, \quad   \left\{q^{a\,\alpha }, q^{b\, \beta}\right\} = 4   \imath \epsilon^{ab } \epsilon^{\alpha \beta} \, {\cal H}, \qquad
\left\{Q^{i\,A} , q^{a\,\alpha} \right\} = 0. 
\ee
to superconformal ones. Thus, we need the generators of dilatation ${\cal D}$, conformal boost ${\cal K}$ and superconformal transformations $S^{i\,A}, s^{a\,\alpha}$ defined as follows\footnote{The basic brackets for the bosonic and fermionic fields are listed in the Appendix.}:
\be\label{conf} 
{\cal D}= \frac{1}{2} r\,p_r, \; {\cal K}= \frac{1}{2} r^2, \quad S^{i\,A} = r \phi^{i\,A} , \; 
s^{a\,\alpha} = r \chi^{a\, \alpha} .
\ee
The basic Poisson brackets have the form
\be\label{N8Conf}
\left\{S^{ i\,A} , S^{j\, B} \right\} = 4 \imath\epsilon^{ij} \epsilon^{AB} {\cal K}, \quad   \left\{s^{a\,\alpha }, s^{b\, \beta}\right\} = 4   \imath \epsilon^{a b } \epsilon^{\alpha \beta} \, {\cal K}, \qquad
\left\{S^{i\,A} , s^{a\,\alpha} \right\} = 0. 
\ee
Depending on the structure of superconformal symmetry, the generators of $R$-symmetry will
include the bosonic  $\{J^{ij}, W^{AB}, X^{ab}, Y^{\alpha\beta}, V^{i\,A\,a\,\alpha}\}$ \p{bosonicso8} and/or fermionic  $so(8)$ generators $\{\hJ^{ij}, \hW^{AB}, \hX^{ab}, \hY^{\alpha\beta}, \hV^{i\,A\,a\,\alpha}\}$ \p{fsu2},\p{fV}  or some subset of these generators.

The idea of our Ansatz for the $\cN=8$ supercharges comes from the  papers  \cite{hkn1,tigran} 
in which the supercharges have the form \p{01}
$$
Q^{i\,A} =p_r\phi^{i\,A} + \frac{\Theta^{i\,A}}{r},\qquad \qquad  q^{a\, \alpha} = 
p_r\chi^{a\, \alpha}+\frac{ \Upsilon^{a\, \alpha}}{r} .
\label{sc}
$$
The composites 
$\Theta^{i\, A}$ and  $\Upsilon^{a\, \alpha}$ have the structure "$(R-symmetry\; generators) \times \, fermions$". Thus, the terms cubic in  fermions in supercharges correspond to fermionic realizations of $R$-symmetry generators while terms linear in fermions in supercharges correspond to bosonic $R$-symmetry generators.

\setcounter{equation}0
\section{$Osp(8|2)$ superconformal mechanics}
From the beginning we have at hand two types of fermions,  $\phi^{i\, A}$ and $\chi^{a\, \alpha}$.
They possess the indices of the fundamental representations of four different $SU(2)$ groups. If we want to have the full $SO(8)$ group as $R$-symmetry,  these $su(2)$ groups have to be unbroken. Therefore, the unique admissible three-linear cross fermionic terms in the supercharges \p{01} read:
\be
\left( \phi^{j\, B}\phi_{j\, B} \right)\phi^{i\, A}\; \mbox{and} \;\left( \chi^{a\, \alpha} \chi_{a\, \alpha}\right) \phi^{i\, A} \; \mbox{ in } \; \Theta^{i\, A}\quad \mbox{and} \qquad
\left( \chi^{b\, \beta} \chi_{b\, \beta}\right) \chi^{a\,\alpha} \; \mbox{and} \; \left( \phi^{i\, A}\phi_{i\, A} \right) \chi^{a\, \alpha}\;  \mbox{ in }\; \Upsilon^{a\, \alpha}.
\ee
However, these terms are equal to zero because in virtue of the fermionic structure we have
\be
\left( \chi^{a\, \alpha} \chi_{a\, \alpha}\right) =0 \quad \mbox{and} \quad
\left( \phi^{i\, A}\phi_{i\, A} \right) =0.
\ee

Now we can include the possible bosonic $R$-symmetry terms preserving all $so(8)$ symmetries as
\bea\label{sco8}
Q^{i\, A} & =&  p_r \phi^{i\, A} +\frac{1}{r}\left( \gamma_1 \cJ^{ij}\phi_j{}^A+ \gamma_2 \cW^{AB}\phi^i{}_B+ \gamma_3 \cV^{i\,A\, a\,\alpha} \chi_{a\, \alpha} \right), \nn \\
q^{a\,\alpha}& = & p \chi^{a\, \alpha}+\frac{1}{r}\left( \rho_1 \cX^{ab}\chi_b{}^\alpha+ \rho_2 \cY^{\alpha\beta} \chi^a{}_\beta+ \rho_3 \cV^{i\,A\, a\,\alpha} \phi_{i\, A}\right).
\eea

To obey  $\cN=8$ super-Poincar\'e algebra \p{N8P}, the parameters $\gamma_{1-3},\rho_{1-3}$ have to be uniquely fixed as
\be\label{sol1}
\gamma_{1}=\gamma_2=\rho_1=\rho_2=1, \quad \gamma_3 = -\frac{1}{2}, \; \rho_3= \frac{1}{2} .
\ee
Thus, the proper supercharges read
\bea\label{scosp}
Q^{i\, A} & =&  p_r \phi^{i\, A} +\frac{1}{r}\left( \cJ^{ij}\phi_j{}^A+  \cW^{AB}\phi^i{}_B - \frac{1}{2} \cV^{i\,A\, a\,\alpha} \chi_{a\, \alpha}, \right), \nn \\
q^{a\,\alpha}& = & p_r \chi^{a\, \alpha}+\frac{1}{r}\left( \cX^{ab}\chi_b{}^\alpha+  \cY^{\alpha\beta} \chi^a{}_\beta+ \frac{1}{2} \cV^{i\,A\, a\,\alpha} \phi_{i\, A}\right).
\eea
The corresponding Hamiltonian has the form
\be\label{hso8}
H_{Osp(8|2)} = \frac{1}{2} p_r^2 +\frac{1}{r^2} \left( -\frac{1}{2} {\cal C}_{so(8)} +\frac{5}{8} ({\cal C}_{so(8)}|_{fermions \rightarrow 0})\right) .
\ee 
The $so(8)$ Casimir operator $\cC_{so(8)}$ is defined in a standard way as
\be 
{\cal C}_{so(8)} \equiv \cJ^{ij} \cJ_{ij} +\cW^{AB} \cW_{AB} +\cX^{ab} \cX_{ab }+\cY^{\alpha\beta} \cY_{\alpha\beta}+\frac{1}{4} \cV^{i\, A\, a\, \alpha } \cV_{i\, A\, a\, \alpha} .
\ee

The generators of  dynamical $SO(8)$ $R$-symmetry of the system appear in the brackets between
the Poincar\'e \p{scosp} and conformal supersymmetry generators \p{conf}:
\bea\label{spconf}
\left\{Q^{i\,A}, S^{j\, B}\right\} & = & 2\, \im\, \epsilon^{ij}\cW^{AB} + 2\, \im\, \epsilon^{AB}  \cJ^{ij} + 4\, \im\, \epsilon^{ij}\epsilon^{AB} D, \nn\\
\left\{Q^{i\,A}, s^{a\, \alpha}\right\} & = & -\, \im\, \cV^{i\,A\,a\,\alpha}, \quad  
\left\{q^{a\, \alpha}, S^{i\, A}\right\}  =  \im\, \cV^{i\,A\, a\, \alpha},   \nn \\
\left\{q^{a\,\alpha}, s^{b\, \beta}\right\} & = & 2\, \im\, \epsilon^{ab} \cY^{\alpha\beta} + 2 \, \im \, \epsilon^{\alpha\beta} \cX^{ab} +
4\, \im\,\epsilon^{ij}\epsilon^{\alpha \beta} D .
\eea

The bosonic part of the Hamiltonian \p{hso8} reads\footnote{The $so(8)$ operators $\ell_{\mu,\nu}$
	are defined in \p{ell}. Note, the term $ \sum_{\mu,\nu=1}^8 \ell_{\mu,\nu} \ell_{\mu,\nu} $ does not fix the norm  $y_\mu y_\mu$, which therefore can be identified with $r^2$, reducing "evident" nine bosonic coordinates to eight.} 
\be\label{ospbos}
H_{bos} =  \frac{1}{2} p_r^2 +\frac{1}{8 r^2}  {\cal C}_{so(8)}|_{fermions \rightarrow 0} =
 \frac{1}{2} p^2 +\frac{1}{16 r^2} \sum_{\mu,\nu=1}^8 \ell_{\mu,\nu} \ell_{\mu,\nu}  .
\ee 

As a consequence of the consideration in section 6, we come to the conclusion that we  deal with a system with eight bosonic and eight fermionic variables, i.e. with the $\cN=8$ supermultiplet  {\bf (8,8,0)}. 

\setcounter{equation}0
\section{${\mathfrak F}(4)$ and $Osp(4^\star|4)$ superconformal mechanics}
One of the possibilities to break $SO(8)$ $R$-symmetry and therefore admit more terms in  supercharges  is to identify, for example, the indices $i,j$ with the indices $\alpha,\beta$. In other words, such identification  breaks $su(2) \times su(2)$ algebra spanned by the generators $J^{ij}$ and $Y^{\alpha\beta}$ down to diagonal $su(2)$ algebra  with the generators $J^{ij} + Y^{ij}$.
The proper Ansatz for the supercharges now reads
\bea\label{1stepdown}
Q^{i\,A}& = & p_r \phi^{i\,A} +\frac{1}{r} \left[\left( m_1 \left(\hJ^{i\,j}+\hY^{i\,j}\right) +m_2 \left( J^{i\, j} + Y^{i\,j}\right)\right) \phi_j{}^A+ \left( m_3 \hW^{AB} +m_4 W^{AB}\right) \phi^i{}_B + \right. \nn \\
&& \left. +\left( m_5 \hV^{i\,A\,a\,j}+ m_6 \hV^{j\,A\,a \,i}+
 m_7 V^{i\,A\,a\, j}+m_8 V^{j\,A\,a\, i}\right)\chi_{a j}\right], \nn \\
q^{a\,i}& = & p_r \chi^{a\,i} +\frac{1}{r} \left[\left( n_1 \left(\hJ^{i\,j}+ \hY^{i\,j}\right)+
 n_2 \left(J^{i\,j}+Y^{i\,j}\right)\right) \chi^a{}_j+ \left(n_3 \hX^{a\,b}+n_4 X^{a\,b}\right) \chi_b{}^i + \right. \nn \\
&& \left. +\left( n_5 \hV^{i\,A\,a\,j} + n_6 \hV^{j\,A\,a\,i}+
 n_7 V^{i\,A\,a\,j}+n_8\hV^{j\,A\,a\,i}\right)\phi_{j A}\right].
\eea
The simplest calculations show that there are several sets of  parameters $m_k,n_k$ to obey
the algebra \p{N8P}.
\subsection{${\mathfrak F}(4)$ supersymmetric mechanics}
The first solution is described by the following supercharges:
\bea\label{F4}
&& Q^{i\,A} =  p_r \phi^{i\,A} -\frac{1}{r} \left[ \frac{2}{3} \left(\hJ^{i\,j}+\hY^{i\,j} -J^{i\, j}- Y^{i\,j}\right) \phi_j{}^A +\frac{4}{9}\left(  \hW^{AB} -3  W^{AB}\right) \phi^i{}_B + \right.  \left. \frac{1}{3}\left( \cV^{i\,A\,a\,j}+\cV^{j\,A\,a \,i}\right)\chi_{a j}\right], \nn \\
&& q^{a\,i} =  p_r \chi^{a\,i} -\frac{1}{r} \left[\frac{2}{3}\left( \hJ^{i\,j}+ \hY^{i\,j} -
J^{i\,j}-Y^{i\,j}\right) \chi^a{}_j +\frac{4}{9} \left(\hX^{a\,b} -3 X^{a\,b}\right) \chi_b{}^i -  \frac{1}{3}\left( \cV^{i\,A\,a\,j}+\cV^{j\,A\,a\,i}\right)\phi_{j A}\right],
\eea
and by the Hamiltonian
\be\label{HF4}
H =\frac{1}{2} p_r^2+ \frac{1}{r^2}\left[ -\frac{1}{3}{\cal C}_{so(7)}+\frac{4}{9} \left({\cal C}_{so(7)}|_{fermions\rightarrow 0}\right)+\frac{2}{9} \left({\cal C}_{so(7)}|_{bosons\rightarrow 0}\right) \right]
.
\ee
Here $\cC_{so(7)}$ is the Casimir operator commuting with the generators $\{\cJ^{ij}+\cY^{ij},\cX^{ab},\cW^{AB}, \cV^{i\,A\,a\,j}+ \cV^{j\,A\,a\,i}\}$:
\be\label{casso7}
{\cal C}_{so(7)} =\left( \cJ^{ij}+\cY^{ij}\right)\left(\cJ_{ij}+\cY_{ij}\right)+ 2 \cX^{ab} \cX_{ab} +
2 \cW^{AB} \cW_{AB} +\frac{1}{8} \left( \cV^{i\,A\,a\,j}+ \cV^{j\,A\,a\,i}\right)
\left( \cV_{i\,A\,a\,j}+ \cV_{j\,A\,a\,i}\right).
\ee

To understand the full dynamical symmetry of the system, one has to calculate the Poisson brackets between
the Poincar\'e \p{F4} and conformal supersymmetry generators \p{conf} (with the indices $\alpha,\beta$ replaced by the indices $i,j$):
\bea\label{F4conf}
\left\{Q^{i\,A}, S^{j\, B}\right\} & = & \frac{8}{3}\, \im\, \epsilon^{ij}\cW^{AB}+ \frac{4}{3}\, \im\, \epsilon^{AB} \left( \cJ^{ij}+\cY^{ij}\right) + 4\, \im\, \epsilon^{ij}\epsilon^{AB} D, \nn\\
\left\{Q^{i\,A}, s^{a\,j}\right\} & = & -\frac{2}{3}\, \im\, \left( \cV^{i\,A\,a\,j}+
\cV^{j\,A\,a\,i}\right), \quad  
\left\{S^{i\, A}, q^{a\, j}\right\}  = \frac{2}{3}\, \im\, \left(\cV^{i\,A\,a\,j}+
\cV^{j\,A\,a\,i}\right),   \nn \\
\left\{q^{a\,i}, s^{b\,j}\right\} & = & \frac{8}{3}\, \im\, \epsilon^{ij} \cX^{a\,b}+ \frac{4}{3} \, \im \, \epsilon^{ab} \left( \cJ^{ij}+\cY^{ij}\right) +
4\, \im\,\epsilon^{ij}\epsilon^{ab} D .
\eea

Thus, we see that instead of $so(8)$ symmetry we now have its subalgebra, spanned by the  generators
\be\label{so7}
\{ \cJ^{ij}+\cY^{ij}, \cX^{ab}, \cW^{AB},  \cV^{i\,A\,a\,j}+ \cV^{j\,A\,a\,i} \}.
\ee
These 21 generators span $so(7)$ subalgebra in $so(8)$. The crucial Poisson brackets read
\bea
\{\cV^{i\,A\,a\,j}+\cV^{j\,A\,a\,i},\cV^{k\,B\,b\,l}+\cV^{l\,B\,b\,k}\} & = & -8 \epsilon^{ab}\epsilon^{AB} \left( \epsilon^{ik} \left(\cJ^{jl} +\cY^{jl}\right)+
\epsilon^{jl} \left(\cJ^{ik} +\cY^{ik}\right) \right) -  \\
&& 8 \epsilon^{AB} \left( \epsilon^{ik}\epsilon^{jl}+\epsilon^{il}\epsilon^{jk}\right)\cX^{ab}-
8 \epsilon^{ab} \left( \epsilon^{ik}\epsilon^{jl}+\epsilon^{il}\epsilon^{jk}\right)\cW^{AB}. \nn
\eea

Thus, we deal with ${\mathfrak F}(4)$ superconformal symmetry and the operator $\cC_{so(7)}$ in \p{casso7} is the Casimir of $so(7)$ algebra. Note, the pure fermionic parts of the supercharges in \p{F4} coincide with those obtained in \cite{F4scm}. 

The bosonic part of the Hamiltonian \p{HF4} reads 
\be\label{f4bos}
H_{bos} =  \frac{1}{2} p_r^2 +\frac{1}{9 r^2}  \cC_{so(7)}|_{fermions \rightarrow 0} =
\frac{1}{2} p^2 +\frac{1}{9 r^2} \sum_{\mu,\nu=1}^7 \ell_{\mu,\nu} \ell_{\mu,\nu}  .
\ee 

As a consequence of the consideration in {\sl Section 6}, we come to the conclusion that we are dealing with a system with seven bosonic and eight fermionic variables, i.e. with the $\cN=8$ supermultiplet {\bf (7,8,1)}.

Another possibility to include more bosonic fields in  ${\mathfrak F}(4)$
superconformal mechanics is to use the multi-superfields approach developed in \cite{fed2}.

Note, to obtain the version of ${\mathfrak F}(4)$ superconformal mechanics with $\bf (1,8,7)$ supermultiplet, constructed in  \cite{F4scm, toppan3} it is enough to put all bosonic currents 
to zero in the supercharges \p{F4} and Hamiltonian \p{HF4}.

\subsection{$OSp(4^\star |4)$ supersymmetric mechanics with $SO(5) \times SU(2)$ dynamical $R$-symmetry}
If instead of the generators $\cV^{i\,A\,a\,j}+ \cV^{j\,A\,a\,i}$ we  consider the generators
$\cU^{A\,a}$ defined as 
\be
\cU^{A\,a} \equiv \epsilon_{ij} \cV^{i\,A\,a\,j},
\ee
one can immediately check that the four generators $\cU^{A\,a}$ form $so(5)$ algebra together
with the generators $\cX^{ab}$ and $\cW^{AB}$. The crucial Poisson brackets read
\be
\{\cU^{A\,a},\cU^{B\,b} \} =- 8 \epsilon^{AB} \cX^{ab} -8 \epsilon^{ab} \cW^{AB}.
\ee
Note, these generators $\{\cU^{A\,a},\cX^{ab},\cW^{AB}\}$ commute with the $su(2)$ generators
$\cJ^{ij} +\cY^{ij}$. Thus, we have the algebra $so(5) \times su(2)$  which is $R$-symmetry of $osp(4^\star |4)$ superconformal algebra. 

To construct the supercharges obeying the $\cN=8$ super Poincar\'e algebra \p{N8P}, we have two
possibilities.

\subsubsection{\bf Variant with bosonic $so(5)$}

The first solution is described by the following supercharges
\bea\label{scso51}
Q^{i\,A} & = & p_r \phi^{i\,A} +\frac{1}{r} \left[ 2 \left( \hJ^{ij}+\hY^{ij}\right) \phi_j{}^A+
4 \cW^{AB} \phi^i{}_B+ \cU^{A\,a} \chi_a{}^i \right], \nn \\
q^{a\,i} & = & p_r \chi^{a\,i} +\frac{1}{r} \left[ 2 \left( \hJ^{ij}+\hY^{ij}\right) \chi^a{}_j+
4 \cX^{ab} \chi_b{}^i+ \cU^{B\,a} \phi^i{}_B\right].
\eea
The corresponding Hamiltonian has the form
\be\label{hso51}
H =\frac{1}{2} p_r^2+ \frac{1}{r^2} \left[ -2 {\cal C}_{so(5)}+4 \cC_{so(5)}|_{fermions\rightarrow 0}+
 {\cal C}_{so(5)}|_{bosons\rightarrow 0} \right] ,
\ee
where the $so(5)$ Casimir operator is defined in a standard manner as
\be
{\cal C}_{so(5)} =  \cX^{ab} \cX_{ab} +\cW^{AB}\cW_{AB}+\frac{1}{8} \cU^{A\,a} \cU_{A\,a} .
\ee
Note, with our definitions we have the following equality:
\be
\left(\hJ^{ij}+\hY^{ij}\right)\left(\hJ_{ij}+\hY_{ij}\right) =-  {\cal C}_{so(5)}|_{bosons\rightarrow 0}.
\ee

The bosonic part of the Hamiltonian \p{hso51} can be represented as
\be\label{sobos}
H_{bos} = \frac{1}{2} p_r^2 +\frac{2}{r^2}{\cal C}_{so(5)}|_{fermions\rightarrow 0} =
 \frac{1}{2} p_r^2 + \frac{1}{r^2} \sum_{\mu,\nu=1 \neq 5,6,7}^8 \ell_{\mu,\nu} \ell_{\mu,\nu} .
\ee

As a consequence of the consideration in {\sl Section 6} we come to the conclusion that we are dealing with a system with five bosonic and eight fermionic variables, i.e. with the $\cN=8$ supermultiplet {\bf (5,8,3)}.

Finally, the Poisson brackets between the Poincar\'e \p{scso51} and conformal supersymmetry generators \p{conf} read:
\bea\label{so51conf}
\left\{Q^{i\,A}, S^{j\, B}\right\} & = & 8\, \im\, \epsilon^{ij}\cW^{AB}+ 4\, \im\, \epsilon^{AB} \left( \hJ^{ij}+\hY^{ij}\right) + 4\, \im\, \epsilon^{ij}\epsilon^{AB} D, \nn\\
\left\{Q^{i\,A}, s^{a\,j}\right\} & = & 2\, \im\, \epsilon^{ij} \cU^{A\,a}, \quad  
\left\{S^{i\, A}, q^{a\, j}\right\}  = -2 \, \im\, \epsilon^{ij} \cU^{A\,a},   \nn \\
\left\{q^{a\,i}, s^{b\,j}\right\} & = & 8\, \im\, \epsilon^{ij} \cX^{a\,b} -4 \, \im \, \epsilon^{ab} \left( \hJ^{ij}+\hY^{ij}\right) + 4\, \im\,\epsilon^{ij}\epsilon^{ab} D .
\eea

\subsubsection{Variant with bosonic $su(2)$}

The second solution is described by the following supercharges:
\be\label{scso52}
Q^{i\,A} =  p_r \phi^{i\,A} -\frac{2}{r} \left( \cJ^{i\,j}+\cY^{i\,j}\right) \phi_j{}^A, \quad
q^{i\,a} =  p_r \chi^{i\,a} -\frac{2}{r} \left( \cJ^{i\,j}+\cY^{i\,j}\right) \chi_j{}^a
\ee
and by the Hamiltonian
\be\label{hso52}
H =\frac{1}{2} p_r^2+ \frac{1}{r^2}  {\cal C}_{su(2)} ,
\ee
where the $su(2)$ Casimir operator has the standard form
\be
{\cal C}_{su(2)} = \left( \cJ^{ij} +  \cY^{ij}\right) \left( \cJ_{ij}+ \cY_{ij}\right).
\ee

The bosonic part of the Hamiltonian \p{hso52} can be represented as
\be\label{su2bos}
H_{bos} = \frac{1}{2} p_r^2 + \frac{1}{r^2} {\cal C}_{su(2)}|_{fermions \rightarrow 0} =
\frac{1}{2} p_r^2 + \frac{1}{r^2} \sum_{\mu,\nu=5}^7 \ell_{\mu,\nu} \ell_{\mu,\nu} .
\ee

As a consequence of the consideration in {\sl Section 6}, we come to the conclusion that we are dealing with a system with three bosonic and eight fermionic variables, i.e. with the $\cN=8$ supermultiplet {\bf (3,8,5)}.

Now  the Poisson brackets between the Poincar\'e \p{scso52} and conformal supersymmetry generators \p{conf} read
\bea\label{so52conf}
\left\{Q^{i\,A}, S^{j\, B}\right\} & = & 8\, \im\, \epsilon^{ij}\hW^{AB} - 4\, \im\, \epsilon^{AB} \left( \cJ^{ij}+\cY^{ij}\right) + 4\, \im\, \epsilon^{ij}\epsilon^{AB} D, \nn\\
\left\{Q^{i\,A}, s^{a\,j}\right\} & = & 2\, \im\, \epsilon^{ij}  \hU^{A\,a }, \quad  
\left\{S^{i\, A}, q^{a\,j}\right\}  =  -2\, \im\, \epsilon^{ij}  \hU^{A\,a},   \nn \\
\left\{q^{a\,i}, s^{b\,j}\right\} & = & 8\, \im\, \epsilon^{ij} \hX^{ab} - 4 \, \im \, \epsilon^{ab} \left( \cJ^{ij}+\cY^{ij}\right) +
4\, \im\,\epsilon^{ij}\epsilon^{a b} D .
\eea

Thus, we have the same $so(5)$ symmetry generated now by purely fermionic generators
$\{\hU^{A\,a}, \hW^{AB}, \hX^{ab}\}$ while commuting with this symmetry
$su(2)$ algebra acquires the bosonic extension
\be
su(2) \propto \{\cJ^{ij}+\cX^{ij}\}.
\ee

\setcounter{equation}0
\section{$SU(1,1|4)$ supersymmetric mechanics with $su(4) \times u(1)$ dynamical $R$-symmetry}
The last possibility is to identify the pairs of indices:
\be
\{i,j\} \; \mbox{with} \; \{\alpha,\beta\} \qquad \mbox{and} \qquad \{a,b\} \; \mbox{with} \; \{A,B\}.
\ee
The very restrictive conditions are that  only sum of generators $\cJ^{ij}+\cY^{ij}$ and
$\cX^{ab}+\cW^{ab}$  should appear  in the Poisson  brackets of $\cN=8$ supersymmetry and the superconformal generators \p{conf} almost completely fix the form of the supercharges. Moreover, to have  $su(4) \times u(1)$ symmetry, the generators $\cV^{i\,a\,b\,j}$ have to appear in the supercharges in the following combinations:
\be\label{Z}
\cZ^{i\,a\,b\,j} \equiv \cV^{i\,a\,b\,j}+\cV^{i\,b\,a\,j}+\cV^{j\,a\,b\,i}+\cV^{j\,b\,a\,i}.
\ee
One can check that the nine generators $\cZ^{i\,a\,b\,j}$ together with the generators $\cJ^{ij}+\cY^{ij}$ and $\cX^{ab}+\cW^{ab}$ form $su(4)$ algebra:
\bea\label{su4}
\{ \cZ^{i\,a\,b\,j},\cZ^{k\,c\,d\,l}\} = -8 \left(\epsilon^{ac} \epsilon^{bd}+\epsilon^{bc}\epsilon^{ad}\right)\left( \epsilon^{ik}(\cJ^{jl}+\cY^{jl})+
\epsilon^{jk}(\cJ^{il}+\cY^{il})+ \epsilon^{il}(\cJ^{jk}+\cY^{jk})+
\epsilon^{jl}(\cJ^{ik}+\cY^{ik})\right)- &&  \nn \\
8 \left(\epsilon^{ik} \epsilon^{jl}+\epsilon^{jk}\epsilon^{il}\right)\left( \epsilon^{ac}(\cX^{bd}+\cW^{bd})+
\epsilon^{bc}(\cX^{ad}+\cW^{ad})+ \epsilon^{ad}(\cX^{bc}+\cW^{bc})+
\epsilon^{bd}(\cX^{ac}+\cW^{ac})\right). && \nn
\eea
Finally, the $U(1)$ generator can be defined as
\be\label{U}
\cU = \cU^i{}_i = \cV^{i\,a}{}_{a\,i}, \qquad U = V^{i\,a}{}_{a\,i},\;\hU = \hV^{i\,a}{}_{a\,i}.
\ee

To construct the $\cN=8$ supercharges, we again have two possibilities.
\subsection{Variant with bosonic $su(4)$}

The first solution is described by the following supercharges:
\bea\label{scsu41}
Q^{i\,a} & = & p_r \phi^{i\,a} +\frac{1}{r} \left[ \left( \cJ^{ij}+\cY^{ij}\right) \phi_j{}^a+
\left( \cX^{ab}+\cW^{ab}\right) \phi^i{}_b -\frac{1}{4}\left( \cZ^{i\,a\,b\,j}+3 \epsilon^{ab}\epsilon^{ij} \hU \right) \chi_{bj} \right], \nn \\
q^{a\,i} & = & p_r \chi^{a\,i} +\frac{1}{r} \left[\left( \cJ^{ij}+\cY^{ij}\right) \chi^a{}_j+
\left(\cX^{ab}+\cW^{ab}\right) \chi_b{}^i+ \frac{1}{4}\left( \cZ^{i\,a\,b\,j}+3 \epsilon^{ij} \epsilon^{ab} \hU\right)  \phi_{jb}\right].
\eea
The corresponding Hamiltonian has the form
\be\label{hsu41}
H =\frac{1}{2} p_r^2+ \frac{1}{r^2} \left[ -\frac{1}{2} {\cal C}_{su(4)}+\frac{3}{4} {\cal C}_{su(4)}|_{fermions\rightarrow 0}+ \frac{3}{10}
{\cal C}_{su(4)}|_{bosons\rightarrow 0} \right] ,
\ee
where the $su(4)$ Casimir operator is defined in a standard manner as
\be
{\cal C}_{su(4)} = (\cJ^{ij}+\cY^{ij})(\cJ_{ij}+\cY_{ij})+ (\cX^{ab}+\cW^{ab})(\cX_{ab}+\cW_{ab})+ \frac{1}{32} \cZ^{i\,a\,b\,j} \cZ_{i\,a\,b\,j} .
\ee
Note, with our parameterization the $U(1)$ fermionic Casimir operator $\hU\,\hU$ is related to
the fermionic part of the $su(4)$ Casimir operator ${\cal C}_{su(4)}|_{bosons\rightarrow 0} $ as
\be
\hU\, \hU =  -\frac{8}{5} {\cal C}_{su(4)}|_{bosons\rightarrow 0} .
\ee

The bosonic part of the Hamiltonian \p{hsu41} can be represented as
\be\label{su4bos}
H_{bos} = \frac{1}{2} p_r^2 + \frac{1}{4 r^2} {\cal C}_{su(4)}|_{fermions \rightarrow 0} =
\frac{1}{2} p_r^2 + \frac{1}{ 4 r^2} \sum_{\mu,\nu=1; \mu,\nu \neq 3}^7 \ell_{\mu,\nu} \ell_{\mu,\nu} .
\ee

As a consequence of the consideration in {\sl Section 6}, we come to the conclusion that we are dealing with a system with six bosonic and eight fermionic variables, i.e. with the $\cN=8$ supermultiplet {\bf (6,8,2)}.

The brackets between the Poincar\'e \p{scsu41} and conformal supersymmetry generators \p{conf} read
\bea\label{su41conf}
\left\{Q^{i\,a}, S^{j\, b}\right\} & = & 2\, \im\, \epsilon^{ij}(\cX^{ab}+\cW^{ab}) +2\, \im\, \epsilon^{ab} \left( \cJ^{ij}+\cY^{ij}\right) + 4\, \im\, \epsilon^{ij}\epsilon^{ab} D, \nn\\
\left\{Q^{i\,a}, s^{b\,j}\right\} & = & -\frac{\im}{2} \cZ^{i\,a\,b\,j}+\frac{\im}{2} \, \epsilon^{ij}\epsilon^{ab}  \hU, \quad  
\left\{S^{i\, a}, q^{b\,j}\right\}  =  \frac{\im}{2} \cZ^{i\,a\,b\,j}-\frac{\im}{2} \, \epsilon^{ij}\epsilon^{ab}  \hU,   \nn \\
\left\{q^{a\,i}, s^{b\,j}\right\} & = & 2\, \im\, \epsilon^{ij}(\cX^{ab}+\cW^{ab})+2 \, \im \, \epsilon^{ab} \left( \cJ^{ij}+\cY^{ij}\right) +
4\, \im\,\epsilon^{ij}\epsilon^{a b} D .
\eea

\subsection{Variant with bosonic $u(1)$}

This simplest variant of $\cN=8$ superconformal mechanics is described by the following supercharges
(the generator $\cU$ was defined in \p{U}):
\be\label{scsu42}
Q^{i\,a} = p_r \phi^{i\,a} +\frac{1}{2 r} \cU \chi^{a\,i}, \quad
q^{a\,i} = p_r \chi^{a\,i} -\frac{1}{2 r} \cU \phi^{i\,a} .
\ee
The corresponding Hamiltonian is also very simple:
\be\label{hsu42}
H = \frac{1}{2} p_r^2 + \frac{1}{8 r^2} \cU\,\cU.
\ee
The bosonic part of the Hamiltonian \p{hsu42} reads
\be\label{u1bos}
H_{bos} =   \frac{1}{2} p_r^2 + \frac{2}{r^2} \ell_{3,8} \ell_{3,8}.
\ee

It  follows from the consideration in {\sl Section 6} that now we deal with a system with two bosonic and eight fermionic variables, i.e. with the $\cN=8$ supermultiplet {\bf (2,8,6)}.

All $su(4)$ generators from $R$-symmetry have the fermionic nature. They all  appear in the
Poisson brackets between the Poincar\'e \p{scsu42} and conformal supersymmetry generators \p{conf}:
\bea\label{su42conf}
\left\{Q^{i\,a}, S^{j\, b}\right\} & = & 2\, \im\, \epsilon^{ij}(\hX^{ab}+\hW^{ab}) +2\, \im\, \epsilon^{ab} \left( \hJ^{ij}+\hY^{ij}\right) + 4\, \im\, \epsilon^{ij}\epsilon^{ab} D, \nn\\
\left\{Q^{i\,a}, s^{b\,j}\right\} & = & -\frac{\im}{2} \hZ^{i\,a\,b\,j}+\im \, \epsilon^{ij}\epsilon^{ab}\left( \cU -\frac{1}{2} \hU\right), \quad  
\left\{S^{i\, a}, q^{b\,j}\right\}  =  \frac{\im}{2} \hZ^{i\,a\,b\,j} -\im \, \epsilon^{ij}\epsilon^{ab}\left( \cU -\frac{1}{2} \hU\right),  \nn \\
\left\{q^{a\,i}, s^{b\,j}\right\} & = & 2\, \im\, \epsilon^{ij}(\hX^{ab}+\hW^{ab})+2 \, \im \, \epsilon^{ab} \left( \hJ^{ij}+\hY^{ij}\right) +
4\, \im\,\epsilon^{ij}\epsilon^{\alpha \beta} D .
\eea
Here the generator $\hZ^{i\,a\,b\,j} $ is the fermionic part of the generator $\cZ$ \p{Z}
\be\label{hZ}
\hZ^{i\,a\,b\,j} \equiv \hV^{i\,a\,b\,j}+\hV^{i\,b\,a\,j}+\hV^{j\,a\,b\,i}+\hV^{j\,b\,a\,i}.
\ee

\setcounter{equation}0
\section{The geometry of configuration space}

In the previous Sections  we considered  superextensions of conformal mechanics on the $2(M+1)$-dimensional phase space equipped  with the canonical Poisson brackets  
\be
\{p_r,r\}=1,\qquad \{p_\mu,y^\nu\}=\delta_\mu^\nu,\qquad \mu,\nu=1,\ldots,M \;.
\ee
We have determined the bosonic parts of the supersymmetric Hamiltonians as 
\be
H=\frac{p^2_r}{2}+\frac{c_0^2\sum_{\mu,\nu=1}^Ml_{\mu,\nu}l_{\mu,\nu} }{r^2},\quad{\rm where}\quad  l_{\mu,\nu}=p_\mu y^\nu-p_\nu y^\mu. 
\label{hb}\ee
The values of the constant $c_0$ and the number of   variables $y^\mu$  for the concrete models proposed in the previous Sections are  as follows \footnote{Previously, the classification of all possible superconformal algebras was carried out in \cite{toppan4}.}:
\begin{enumerate}
	\item $c_0=1/4$, $M=8$ for the $OSp(8|2)$ superconformal mechanics, see  Section 3, Eq.\p{ospbos}
	\item $c_0=1/3$, $M=7$ for the ${\mathfrak F} (4)$ superconformal mechanics, see  Subsection 4.1, Eq.\p{f4bos}
	\item $c_0=1$, $M=5$ for the $OSp(4^\star |4)$ superconformal mechanics  with bosonic $so(5)$ symmetry, see    Subsection 4.2.1, Eq.\p{sobos}
	\item $c_0=1$, $M=3$ for the $OSp(4^\star |4)$ superconformal mechanics  with bosonic $su(2)=so(3)$ symmetry, see Subsection 4.2.2, Eq.\p{su2bos}
	\item $c_0=1/2$, $M=6$ for the $SU(1,1|4)$ superconformal mechanics  with bosonic $su(4)=so(6)$ symmetry, see Subsection 5.1, Eq.\p{su4bos}
	\item $c_0=1$, $M=2$ for the $SU(1,1|4)$ superconformal mechanics  with bosonic $u(1)$ symmetry, see   Subsection 5.2, Eq.\p{u1bos}
	\item $c_0=0$ for all superconformal mechanics except for  $OSp(8|2)$ superconformal mechanics. This case corresponds to the $\bf (1,8,7)$ supermultiplet, and it is achieved by putting to zero  all bosonic currents entering into the supercharges.
\end{enumerate}

Let  us  perform   the  canonical transformation  
$(y^\mu, p_\mu,r,p_r,)\to (x^\mu,\pi_\mu,  R, p_R)$ with 
\be
R=\sqrt{\sum_{\nu=1 }^M  y^\nu y^\nu}\;:=|\mathbf{y}|, \quad 
p_R=\frac{\sum_{\mu=1 }^M  p_\mu y^\mu}{|\mathbf{y}|}, \quad x^\mu= {r} \frac{y^\mu}{|\mathbf{y}|},\quad \pi_\mu=
\frac{ |\mathbf{y}|}{r}p_\mu +  \left(p_r-\frac{\sum_{\lambda=1  }^M p_\lambda y^\lambda}{r}\right) \frac{y^\mu}{|\mathbf{y}| }
,
\ee
and get the Poisson brackets   defined by the relations
$
\{p_R, R\}=1$, $ \{\pi_\mu, x^\nu\}=\delta_\mu^\nu$.

In these terms the   Hamiltonian \eqref{hb} reads
\be
{H}_{bos}= 2 c^2_0  \sum_{\mu,\nu=1  }^M g^{\mu\nu } \pi_\mu\pi_\nu ,\quad {\rm with}\quad g^{\mu\nu}=\delta^{\mu\nu}-\left(1-\frac{1}{4c^2_0}\right) \frac{x^\mu x^\nu}{\sum_{\lambda=1}^M x^\lambda x^\lambda}.
\label{hr}\ee
Hence the variables $p_R, R$ do not appear in the Hamiltonian, i.e. we deal with the $M$-dimensional system.

From expression   \eqref{hr}  we immediately get the metrics of configuration space:
\be
(ds)^2=\sum_{\mu,\nu =1}^M  g_{\mu\nu}dx^\mu dx^\nu,\qquad g_{\mu\nu}=\delta_{\mu\nu}+ (4c^2_0-1)\frac{x^\mu x^\nu}{ {\sum_{\lambda=1 }^M x^\lambda x^\lambda}} ,\qquad\sum_{\lambda=1}^M g^{\mu\lambda}g_{\lambda\nu}=\delta^\mu_\nu.
\label{me}\ee
Equivalently, 
\be
(ds)^2=|\mathbf{\tilde{x}}|^{1/2c_0-1} \sum_{\mu=1}^Md {\tilde{x}}^\mu d {\tilde{x}}^\mu ,\qquad {\rm where}\quad \tilde{x}^\mu=|\mathbf{x}|^{2c_0}\frac{x^\mu}{ {|\mathbf{x}|}},\qquad |\mathbf{x}|:=\sqrt{\sum_{\lambda=1}^M x^\mu x^\mu} ,\quad|\mathbf{\tilde{x}}|:=\sqrt{\sum_{\lambda=1}^M \tilde{x}^\mu \tilde{x}^\mu} .
\ee

For further analyses we  separate the following cases:
\begin{itemize}
	\item {\sl  Cases 1,2}:  $4c^2_0-1<0$, the  $8$-dimensional $Osp(8|2)$ and $7$-dimensional ${\mathfrak F} (4)$ superconformal mechanics.
	
	Their configuration spaces are  the $M=8,7$-dimensional Euclidean cones  in the $(M+1)$-dimensional Minkowski spaces. Namely, the metrics \eqref{me} can be written as 
	
	\be
	(ds)^2= \left(-dx^0dx^0+\sum_{\mu=1}^Md {x}^\mu d{x}^\mu\right)|_{ x^0= \sqrt{1-4c^2_0}|\mathbf x|}.
	\ee
	
	\item {\sl  Cases 3,4}:  $4c^2_0-1=3$, the  $5$-dimensional $Osp(4^\star|4)$ mechanics with bosonic $so(5)$ symmetry and $3$-dimensional $Osp(4^\star|4)$   superconformal mechanics with bosonic $so(3)$ symmetry.
	
	The configuration spaces are  the $M=3,5$-dimensional  cones  in the $(M+1)$-dimensional Euclidean spaces. The metrics \eqref{me} can be written as 
	
	\be
	(ds)^2= \left(dx^0dx^0+\sum_{\mu=1}^Md {x}^\mu d{x}^\mu\right)|_{ x^0= \sqrt{3}|\mathbf x|}.
	\ee
	\item Case 5: $c_0=1/2$,  the $6$-dimensional $SU(1,1|4)$ superconformal mechanics with bosonic $so(6)$ symmetry. Configuration space is the $6$-dimensional Euclidean space. 
	
	\item Case 6: $c_0=1$, the  $2$-dimensional  $SU(1,1|4)$ superconformal mechanics with bosonic $u(1)$ symmetry.
	
	Similarly to  Cases 3,4, the configuration space of the system is the   cone in the $3$-dimensional Euclidean space. However, in contrast with Cases 3,4, 
	this cone   can be immediately identified with   the two-dimensional Euclidean space with the punctured point by introducing  the appropriate complex coordinates
	\be
	z=\frac43\left(\widetilde{x}^i+\im \widetilde{x}^2\right)^{3/4}\quad:\;\Rightarrow \qquad (ds)^2=dzd\bar z.
	\ee 
	Hence the bosonic part of the respective supersymmetric mechanics describes  the free particle in the two-dimensional Euclidean space.
	
	\item Case 7: $c_0=0$. This case describes the dilaton $r$ interacting with the fermions.  No additional bosonic variables are present.
\end{itemize}

\section{Conclusion}
In the present paper we have constructed the supercharges and Hamiltonians for all variants of 
superconformal mechanics associated with the superalgebras $osp(8|2), \mF(4), osp(4^\star |4)$ and
$su(1,1,|4)$. The fermionic and bosonic fields involved were arranged into generators spanning $so(8), so(7), so(5)\times su(2)$ and $su(4) \times u(1)$ $R$-symmetries of the  corresponding superconformal algebras. The bosonic and fermionic parts of these  $R$-symmetry generators   separately define the constants of motion and form the same  algebras. 
The   angular parts of supercharges defining the system have the structure 	``$(R-symmetry\;generators)\, \times\, fermions$'', while the angular part of the Hamiltonian is just a sum of  full  Casimir operators and its
purely bosonic and/or fermionic parts. We explicitly demonstrated that the constructed system describes the specific $\mathcal{N}=8$ supersymmetric extension of  free particles on some cone  embedded in  (pseudo-)Euclidean space with the 	fermionic part that can be interpreted as the spin-orbit coupling terms.

One of the most interesting novelty features of our approach is the treatment of so-called isospin variables, which have been widely used in recent years as the coordinates-momenta in the first order formalism. It would be very intriguing to consider from
this point of view  $\cN=4$ supersymmetric mechanics with isospin variables \cite{brazil, Kon1, Kon2, fil0, Kon3, bk}. Another intriguing  subject concerns the $\cN=4$ variant of the structure 	``$(R-symmetry\; generators)\, \times\, fermions$'' in the supercharges of $\cN=4$ superconformal  mechanics. We expect that this consideration will give rise to new systems with  the spin-orbit coupling terms. But of course, one can save the interpretation of these variables as the isospin 
ones just using the definitions \p{bosonicso8} and the brackets \p{PB1}.

Finally, we would like to stress that the explicit embedding of  the algebras $so(7), \, so(5) \times su(2)$, and $su(4)\times u(1)$ into $so(8)$ we constructed and discussed in this paper
plays an essential role, providing the possibilities to explicitly construct the corresponding supercharges.

{\large{Acknowledgments.}}
The work of A.N. was partially  supported  by the  
Armenian State Committee of Science within the project  21AG-1C062 .

\def\theequation{A.\arabic{equation}}
\setcounter{equation}0
\section*{Appendix}
\noindent{\bf Basic ingredients} \vspace{0.5cm}

To construct the  supercharges spanning $\cN=8$ super Poincar\'{e} algebra, we will first need eight fermions $\{ \phi^{i A}, \chi^{a \alpha}\}$ obeying the following Poisson brackets:
\be\label{fermiPB}
\{ \phi^{i A}, \phi^{j B}  \} = 2\,\im \, \epsilon^{i j} \epsilon^{A B}, \qquad \{ \chi^{a \alpha}, \chi^{b \beta} \}= 2\,\im \, \epsilon^{ab} \epsilon^{\alpha \beta}\;, \qquad i,A,a,\alpha = 1,2
\ee
and possessing the following conjugation properties\footnote{All the $su(2)$ indices are raised and lowered as follows: ${\cal A}^a = \epsilon^{ab}{\cal A}_b, {\cal A}_a = \epsilon_{ab} {\cal A}^b$, where the anti-symmetric tensor $\epsilon_{ab}=-\epsilon^{ba}$ is defined as $\epsilon_{12}=\epsilon^{21}=1$.}:	
\be	
\left( \phi^{i A}\right)^\dagger = \phi_{i A},  \qquad \left( \chi^{a \alpha}\right)^\dagger = \chi_{a \alpha}.
\ee
Note that these fermions form doublets with respect to four $su(2)$ algebras $\hJ^{i j}, \hW^{A B}, \hX^{a b}, \hY^{\alpha \beta}$:
\bea\label{fsu2}
\hJ^{ij} = \frac{\im}{4}  \phi^{i A} \phi^j{}_A  &\quad \Rightarrow \quad & \left\{\hJ^{ij}, \hJ^{kl}\right\} = -\epsilon^{ik} \hJ^{jl} -\epsilon^{jl}\hJ^{ik} , \nn \\
\hW^{A B} = \frac{\im}{4} \phi^{i A} \phi_i{}^B &\quad \Rightarrow \quad & \left\{\hW^{A B}, \hW^{C D}\right\} = -\epsilon^{A C} \hW^{B D} -\epsilon^{B D}\hW^{A C} , \nn \\
\hX^{a b } = \frac{\im}{4} \chi^{a \alpha} \chi^{b}{}_\alpha & \Rightarrow & 
\left\{\hX^{a b}, \hX^{c d}\right\} = -\epsilon^{a c} \hX^{b d} -\epsilon^{b d}\hX^{a c} , \nn \\
\hY^{\alpha \beta}=\frac{\im}{4} \, \chi^{a \alpha} \chi_{a}{}^\beta & \Rightarrow & 
\left\{\hY^{\alpha \beta}, \hY^{\gamma \delta}\right\} = -\epsilon^{\alpha \gamma} \hY^{\beta \delta} -\epsilon^{\beta \delta}\hY^{\alpha \gamma} .
\eea
We also need the generators $\hV^{i A a \alpha}$ from the coset $so(8)/su(2)^4$:
\bea\label{fV}
&& \hV^{i A a \alpha} = \im\, \phi^{i A} \chi^{a \alpha} \; \Rightarrow \nn \\
&& \left\{\hV^{i A a \alpha},\hV^{j B b \beta}\right\} = -4 \left( \epsilon^{AB}\epsilon^{ab} \epsilon^{\alpha \beta} \hJ^{ij} + \epsilon^{ij}\epsilon^{ab} \epsilon^{\alpha \beta} \hW^{AB}+ \epsilon^{ij}\epsilon^{AB} \epsilon^{\alpha \beta} \hX^{ab} + \epsilon^{ij}\epsilon^{AB} \epsilon^{a b} \hY^{\alpha \beta}\right). 
\eea

The bosonic variables include the ``radial'' coordinate $r$ with the momenta $p_r$
\be 
\{ p_r , r \} =1, 
\ee
and sixteen additional bosonic variables $\{B^{a A}_m, Z^{i \alpha}_m\}, \; m=1,2$ obeying the Poisson brackets 
\be\label{PB1}
\left\{ B^{a A}{}_m, B^{b B}{}_n \right\} = 2 \epsilon^{a b} \epsilon^{A B} \epsilon_{m n}, \quad
\left\{ Z^{i \alpha}{}_m, Z^{j \beta}{}_n \right\} = 2 \epsilon^{i j } \epsilon^{\alpha \beta} \epsilon_{m n},
\ee
and possessing the following conjugation properties:
\be	
\left( B^{a A}{}_m\right)^\dagger = B_{a A\, m},  \qquad \left( Z^{i \alpha}{}_m\right)^\dagger = Z_{i \alpha\, m}.
\ee
The bosonic analogues of the $so(8)$ generators can be constructed as\footnote{Note that in contrast with the indices $\{i,A,a,\alpha\}$  of the fundamental representations of four $su(2)$ algebras, the indices $m,n =1,2$ just numerate the number of the fields. However,  the metric in the
	space of these fields is chosen to be $\epsilon^{mn}$ \cite{fed1}. Therefore, any expression of the type ${\cA^m \cA_m}$ means $\epsilon^{mn} \cA_n \cA_m$.} 
\bea
&&J^{ij} =\frac{1}{4} Z^{i \alpha\, m}Z^j{}_{\alpha\,m}, \; W^{AB} =\frac{1}{4} B^{a A\, m}B_a{}^B{}_m, \; X^{a b}= \frac{1}{4} B^{a A\,m}B^b{}_{A\,m}, \; Y^{\alpha\beta}= \frac{1}{4} Z^{i\alpha\,m}Z_i{}^\beta{}_m, \nn \\
&& V^{i\,A\,a\,\alpha}= Z^{i\,\alpha\,m}B^{a\,A}{}_m . \label{bosonicso8}
\eea
The calligraphic letters $\cJ,\cW,\cX,\cY,\cV$ denote the full generators, e.g. $\cJ^{ij} =J^{ij}+\hJ^{ij}, \cW^{AB}=W^{AB}+\hW^{AB}$, etc.\\

\noindent{\bf Different (standard) realization of $so(8)$}
\vspace{0.5cm}

Usually,  variables like $B^{a A}{}_m, Z^{i \alpha}{}_m$ with the non-trivial Poisson brackets
\p{PB1} are associated with the so-called isospin degrees of freedom.
Supersymmetric mechanics describing the motion of an isospin particle in the background non-Abelian gauge fields has attracted much attention in the last decades \cite{brazil, Kon1, Kon2,
	fil0, Kon3, bk}, especially due to their close relation with higher-dimensional Hall effects and their extensions \cite{HE} as well as with the supersymmetric versions
of various Hops maps (see e.g. \cite{brazil}). The key point of any possible construction is to find a proper ``room''  for semi-dynamical ``isospin'' variables that have to be invented
for the description of monopole-type interactions in Lagrangian mechanics. In supersymmetric systems these isospin variables should belong to some supermultiplets containing  additional fermions accompanying the isospin variables. 

However, one can try to treat these variables as additional coordinates and momenta in the first order Lagrangians. The simplest way to introduce these new coordinates and momenta in the supersymmetric mechanics we are dealt with is to associate
the fields $\{B^{a A}{}_m, Z^{i \alpha}{}_m\}$ with lower index $m=2$ with the momenta and the fields with lower index $m=1$ with the coordinates. One of the possible associations reads
\bea
&& B^{11}{}_1=y^1 + \im y^2, \;B^{12}{}_1=y^3 + \im y^4, \;B^{21}{}_1= -y^3 +\im y^4, \; B^{22}{}_1=y^1 - \im y^2,  \nn \\
&& B^{11}{}_2=-p_1 - \im p_2, \;B^{12}{}_2=-p_3-\im p_4, \; B^{21}{}_2 =p_3 -\im p_4, \; B^{22}{}_2 = -p_1+\im p_2, \label{BBb} \\
&& Z^{11}{}_1= y^5+\im y^6, \; Z^{12}{}_1 = y^8+\im y^7, \; \; Z^{21}{}_1=-y^8+\im y^7, \; Z^{22}{}_1=y^5-\im y^6, \nn \\
&& Z^{11}{}_2=-p_5-\im p_6, \; Z^{12}{}_2=-p_8-\im p_7, \; Z^{21}{}_2= p_8-\im p_7, \; Z^{22}{}_2=-p_5+\im p_6 .\label{zz}
\eea
One can check that with the standard brackets
\be
\{p_\mu,y^\nu\} =\delta_\mu^\nu ,\qquad \mu,\nu=1,\ldots, 8
\ee
the brackets \p{PB1} are correctly reproduced.

In terms of the coordinates $y^\mu$ and momenta $p_\mu$, the $so(8)$ generators have the standard form
\be\label{ell}
\ell_{\mu,\nu} = p_\mu y^\nu - p_\nu y^\mu .
\ee
The $so(8)$ generators $J^{ij}, W^{AB},  X^{a b}, Y^{\alpha \beta},V^{i\,A\,a\,\alpha}$ \p{bosonicso8} are related with $\ell_{\mu\nu}$ as follows:
\bea
&& J^{11}= \frac{1}{2}\left( \ell_{5,8}+ \im \ell_{5,7}+\im \ell_{6,8}-\ell_{6,7}\right), \;
J^{12} = -\frac{\im}{2}\left(\ell_{5,6}-\ell_{7,8}\right),\;
J^{22}=\frac{1}{2}\left( \ell_{5,8} - \im \ell_{5,7}-\im \ell_{6,8}-\ell_{6,7}\right), \nn \\
&& W^{11}= \frac{1}{2}\left(- \ell_{1,3} + \im \ell_{1,4}-\im \ell_{2,3}-\ell_{2,4}\right), \;
W^{12} = \frac{\im}{2}\left(-\ell_{1,2}+\ell_{3,4}\right),\;
W^{22}=\frac{1}{2}\left( -\ell_{1,3} - \im \ell_{1,4} + \im \ell_{2,3}-\ell_{2,4}\right), \nn \\
&& X^{11}= \frac{1}{2}\left( \ell_{1,3} + \im \ell_{1,4}+\im \ell_{2,3}-\ell_{2,4}\right), \;
X^{12} = -\frac{\im}{2}\left(\ell_{1,2}+\ell_{3,4}\right),\;
X^{22}=\frac{1}{2}\left( \ell_{1,3} - \im \ell_{1,4}-\im \ell_{2,3}-\ell_{2,4}\right), \nn \\
&& Y^{11}= \frac{1}{2}\left( -\ell_{5,8} + \im \ell_{5,7}-\im \ell_{6,8}-\ell_{6,7}\right), \;
Y^{12} = \frac{\im }{2}\left(\ell_{5,6}+\ell_{7,8}\right),\;
Y^{22}=\frac{1}{2}\left( -\ell_{5,8} - \im \ell_{5,7}+\im \ell_{6,8}-\ell_{6,7}\right), \nn \\
&& V^{1111} = -\ell_{1,5} - \im \ell_{1,6} -\im \ell_{2,5} + \ell_{2,6},\;
V^{1112} = -\ell_{1,8}- \im \ell_{1,7}-\im \ell_{2,8}+\ell_{2,7},\; \nn\\
&& V^{1121} = \ell_{3,5}+ \im \ell_{3,6} -\im \ell_{4,5}+\ell_{4,6},\;
V^{1122} =\ell_{3,8}+ \im \ell_{3,7} -\im \ell_{4,8}-\ell_{4,7},\; \nn\\
&& V^{1211} =- \ell_{3,5} - \im \ell_{3,6}-\im \ell_{4,5}+\ell_{4,6},\;
V^{1212} =-\ell_{3,8} - \im \ell_{3,7} - \im \ell_{4,8}+\ell_{4,7},\; \nn\\
&& V^{1221} = -\ell_{1,5} - \im \ell_{1,6}+\im \ell_{2,5}-\ell_{2,6},\;
V^{1222} = -\ell_{1,8}- \im \ell_{1,7}+\im \ell_{2,8} -\ell_{2,7}, \nn\\
&& V^{2111} = \ell_{1,8}- \im \ell_{1,7}+\im \ell_{2,8}+\ell_{2,7},\;
V^{2112} = -\ell_{1,5} + \im \ell_{1,6}- \im \ell_{2,5}-\ell_{2,6},\; \nn\\
&& V^{2121} =-\ell_{3,8}+ \im \ell_{3,7}+\im \ell_{4,8}+\ell_{4,7},\;
V^{2122} = \ell_{3,5} - \im \ell_{3,6}-\im \ell_{4,5} - \ell_{4,6},\; \nn\\
&& V^{2211} = \ell_{3,8}- \im \ell_{3,7} + \im \ell_{4,8}+\ell_{4,7},\;
V^{2212} = -\ell_{3,5}+ \im \ell_{3,6} - \im \ell_{4,5} -\ell_{4,6}, \nn\\
&& V^{2221} =  \ell_{1,8} - \im \ell_{1,7} -\im \ell_{2,8} - \ell_{2,7},\;
V^{2222} =  -\ell_{1,5} + \im \ell_{1,6} +\im \ell_{2,5}+\ell_{2,6}.
\eea
Note, the expression for the $so(8)$ Casimir operator has the standard form in terms of the generators $\ell_{\mu\nu}$
\be 
{\cal C}_{so(8)} \equiv J^{ij} J_{ij} +W^{AB} W_{AB} +X^{ab}X_{ab }+Y^{\alpha\beta}Y_{\alpha\beta}+\frac{1}{4}V^{i\, A\, a\, \alpha } V_{i\, A\, a\, \alpha} =
\frac{1}{2}\sum_{\mu,\nu=1}^8  \ell_{\mu,\nu} \ell_{\mu, \nu}.
\ee

\end{document}